\documentclass[aps,twocolumn,amssymb,showpacs,floatfix,pra]{revtex4-1}
\usepackage{amssymb}
\usepackage{graphicx}
\usepackage{epsfig}
\usepackage{color}

\begin{document}

\title{Vibrational properties and stability of FePt nanoalloys}

\author{Przemys\l{}aw Piekarz}
\affiliation{Institute of Nuclear Physics, Polish Academy of Sciences,
             Radzikowskiego 152, PL-31342 Krak\'ow, Poland }

\author{Jan \L{}a\.zewski}
\affiliation{Institute of Nuclear Physics, Polish Academy of Sciences,
             Radzikowskiego 152, PL-31342 Krak\'ow, Poland }

\author{Pawe\l{} T. Jochym}
\affiliation{Institute of Nuclear Physics, Polish Academy of Sciences,
             Radzikowskiego 152, PL-31342 Krak\'ow, Poland }

\author{Ma\l{}gorzata Sternik}
\affiliation{Institute of Nuclear Physics, Polish Academy of Sciences,
             Radzikowskiego 152, PL-31342 Krak\'ow, Poland }
             
\author{Krzysztof Parlinski}
\affiliation{Institute of Nuclear Physics, Polish Academy of Sciences,
             Radzikowskiego 152, PL-31342 Krak\'ow, Poland }

\date{\today}

\begin{abstract}
The structural and dynamical properties of FePt nanoparticles were studied within the density functional theory. 
The effect of size and chemical composition on dynamical stability of nanoparticles was investigated for the cuboctahedral
and icosahedral symmetries. In cuboctahedra, the structural distortion is observed, which for systems with odd number of Pt layers leads to lowering of the tetragonal symmetry. 
Significant differences between the vibrational properties of FePt particles and bulk crystal is observed,
but similarly to the FePt crystal, cuboctahedral particles exhibit a strong anisotropy 
of atomic vibrations. The icosahedral particles with perfect shell geometry are unstable due to enlarged distances 
between Fe atoms. They can be stabilized by removing a central atom or replacing
it by a smaller one. The heat capacity and entropy of nanoparticles show typical enhancement due to low-energy
vibrations at the surface layers.
\end{abstract}


\maketitle

\section{Introduction}
Distinct properties of nanoparticles (NPs) stem from
their intermediate size between molecular systems and  
crystals, sharing features with both of these groups.
Nanoparticle alloys (nanoalloys) are bi- or multicomponent metallic particles, 
often with the complex structures and properties, which can be very different 
from those of the corresponding bulk alloys and single-metal nanoparticles
\cite{ferrando,johnston,book}. By adjusting the size and chemical composition,
nanoalloys can be optimized for the applications in catalysis \cite{catal},
nanomedicine \cite{biomed}, and data storage \cite{Sun2}. 

Dynamical properties of nanoobjects are significantly modified comparing 
to bulk systems due to reduced dimensionality and surface effects 
\cite{Bohnen,Frase,Fultz,Kara1,Sun1,Meyer,Kara2,Lazewski1,Slezak,Stankov,Lazewski2,Roldan1,Shafai,Spiridis,Bozyigit,Seiler}. 
The theoretical studies on vibrational properties of isolated NPs, performed 
with the empirical potentials \cite{Kara1,Sun1,Meyer,Kara2}
and density functional theory (DFT) \cite{Roldan1,Shafai},
demonstrated the pronounced changes in the vibrational density of states (VDOS) 
comparing to bulk systems.
The increase of the VDOS at low energies is caused mainly by the surface atoms
with the reduced coordination number, while the shift of the energies 
to higher values above the bulk cut-off is induced by the inner atoms 
with shorter interatomic distances \cite{Kara1,Sun1}.
Vibrational properties of a NP depend substantially on its size and geometry \cite{Shafai},
as well as on the presence of adsorbate atoms \cite{Roldan1}.
In nanoalloys, dynamics of atoms depends additionally on local
environment and atomic coordination \cite{Yildirim}. 
These changes influence thermodynamical properties such as thermal expansion,
heat capacity, melting temperature, and thermal displacements
and make the standard approaches, like Debye model,
not applicable for NPs \cite{Rojas,Shafai}.

Experimentally, the effect of reduced size on vibrational 
properties of NPs have been studied using the nuclear inelastic 
scattering (NIS) \cite{Roldan2,Roldan3,Roldan4} and
the optical methods such as multiple photon dissociation spectroscopy in the far-infrared \cite{Gruene},
time-resolved pump-probe spectroscopy \cite{Juve,Sauceda}, and Raman scattering \cite{Portales,Bayle1,Bayle2,Carles}.
For instance, the energies of the lowest modes (acoustic gap)
and the symmetric modes inducing the radial expansion and contraction of particles (breathing mode) can be
obtained from the pump-probe experiments \cite{Juve,Sauceda}. 

Iron-platinum (FePt) alloys, crystallizing in the layered  $L1_0$ phase,
exhibit an extremely high uniaxial magnetocrystalline anisotropy \cite{Graf}.
FePt is therefore a promising material for applications in 
ultra-dense magnetic recording media provided that well-ordered
$L1_0$-particles of small size can be synthesized \cite{Sun2,Tamada1,Tamada2,Wang}.
Relevant information about the stability and electronic properties of the FePt nanoclusters 
were obtained by theoretical studies \cite{Fortunelli}. 
The Monte Carlo simulations show that melting temperatures of FePt NPs
are much lower than the bulk phase-transition temperature (1572 K) and decrease with the particle size
reduction \cite{Chepulskii}.
The structural order and magnetic properties of FePt nanoalloys with various morphologies
were studied previously within the DFT calculations \cite{Gruner1,Gruner2,Gruner3,Kabir}.
Recently, we have investigated the influence of chemical composition on the stability of FePt 
icosahedra using the {\it ab initio} molecular dynamics (MD) simulations \cite{Jochym}.

The structural anisotropy influences also the dynamical properties of the FePt systems.
The partial Fe VDOS obtained in the composite of FePt NPs \cite{Tamada3},
showed a large difference between the spectra measured parallel and perpendicular to the $c$ axis.
Similar anisotropy was found in the thin films studied experimentally \cite{Couet} and 
by the {\it ab initio} calculations performed for the bulk FePt crystal \cite{Tamada3,Couet,Sternik}.
These studies revealed also differences between the spectra of NPs 
and thin films, mainly at low energies, which result from the surface 
effects and/or phonon confinement.
It has been suggested that the excess of the low-energy modes in the Fe partial DOS
may indicate the appearance of the Fe-terminated surface in the NPs \cite{Couet}.
Also, another interesting question arises: How the dynamical anisotropy depends 
on the size and geometry, and how it affects the thermodynamical properties 
of FePt nanoalloys?

In this paper, we investigate the structural and dynamical properties of selected FePt nanoparticles
with the cuboctahedral and icosahedral geometries using the first-principles DFT methods. 
Comparing to the semi-empirical potentials, the {\it ab initio} approach is limited 
to small nanoparticles, but it explicitly includes the effect of electronic structure 
and magnetic interactions on atomic forces and dynamical properties of particles.
We study the effects of a size and chemical composition on a structural stability
and vibrational spectra of NPs and analyze the structural deformations induced by the soft modes.
Finally, using the vibrational spectra, we calculate the heat capacity and entropy of nanoalloys
as functions of temperature and system size.

The paper is organized as follows. In Sec. II, the calculation method
used in the studies of structural and dynamical properties of FePt NPs
is presented.
Sec. IIIa and IIIb describe the results of calculations for the cuboctahedral 
and icosahedral particles, respectively. In Sec. IV, the thermodynamic properties of nanoparticles
are presented and discussed. Sec.V summarizes presented results.

\section{Calculation method}

The calculations were performed within the DFT as implemented
in the Vienna Ab initio Simulation Package (VASP) \cite{VASP}.
The exchange-correlation functionals are described
by the generalized-gradient approximation (GGA) \cite{GGA}
within the projector augmented-wave method \cite{PAW}.
The calculations were performed in periodic boundary conditions 
in a cubic supercell of properly chosen size $d$ to keep 
the distance between a particle and its images larger than $12$ \AA,
large enough to exclude long-range interactions.
All reciprocal-space summations are performed only at the $\Gamma$ point.
The valence states where optimized with the Fe $3d^64s^2$ and Pt $5d^96s^1$ electron configurations.
The first-order Methfessel-Paxton scheme was applied for the Fermi surface smearing with $\sigma=0.2$~eV.
The cut-off energy of 320 eV was used for the plane-wave expansion.
The convergence criteria were set to $10^{-7}$ eV and $10^{-4}$ eV/\AA\ 
for the total energy and residual forces, respectively.

The calculations for the $L1_0$-FePt crystal with the space group $P4/mmm$ were performed
in the $2\times2\times2$ supercell containing 32 atoms.
The wave functions were sampled over the points generated with Monkhorst--Pack 
scheme using the $4\times4\times4$ {\bf k}--point grid. 
In all calculations we assumed the ferromagnetic (FM)
order on Fe atoms, which is an experimentally observed state 
in the FePt crystal and nanoalloys.

\begin{figure}[t!]
\centering
\includegraphics[width=1\linewidth]{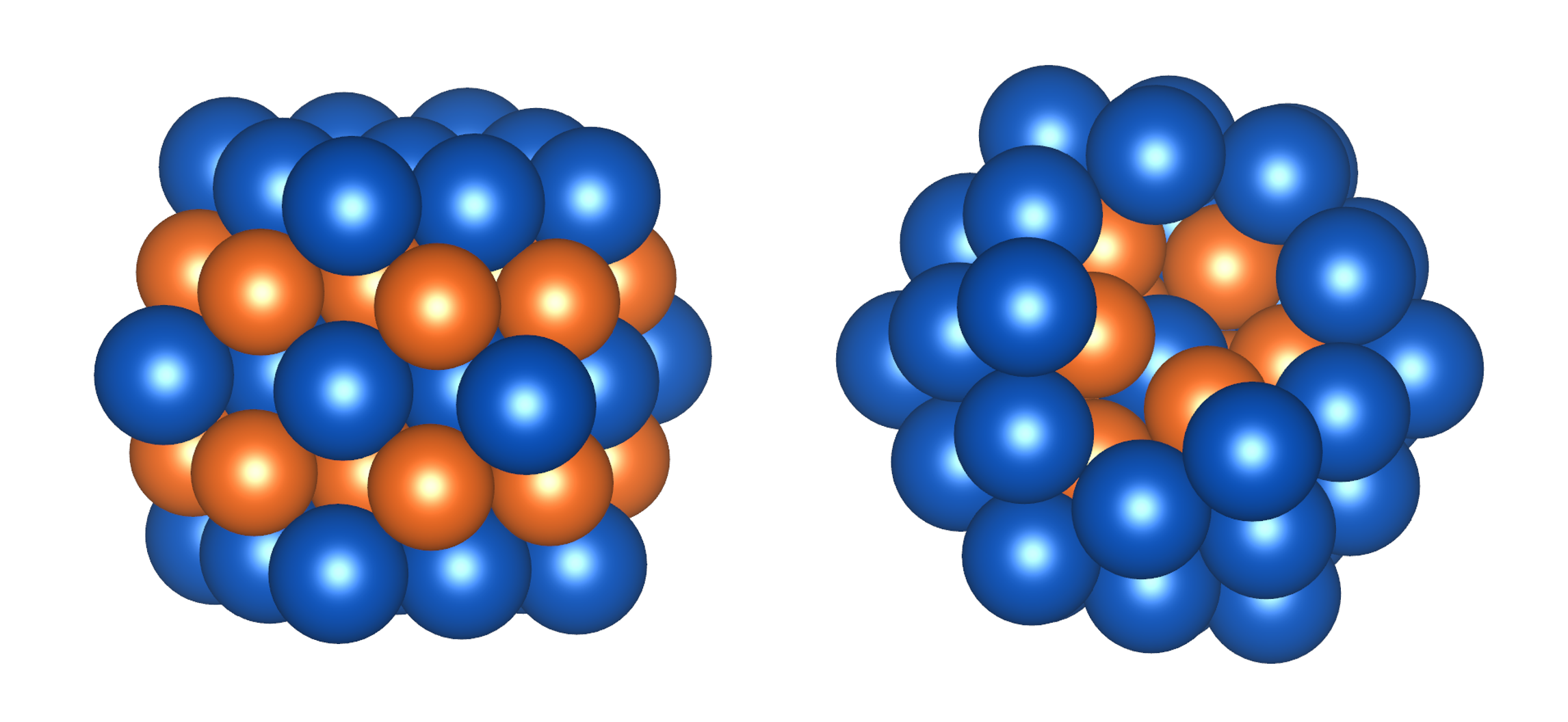}
\caption{Side view of the cuboctahedral Fe$_{24}$Pt$_{31}$ (left)
and icosahedral Fe$_{12}$Pt$_{43}$ (right) nanoparticle. Pt atoms are in blue (dark) and Fe atoms in red (light) colors.
For the icosahedral nanoparticle a few atoms from the outer layers were removed to show its alternating shells structure. The drawing prepared with the VESTA software \cite{VESTA}.}
\label{NPs}
\end{figure}

We have chosen two basic morphologies of NPs: the cuboctahedron with the point group symmetry $D_{4h}$ and the icosahedron ($I_h$) presented in Fig.~\ref{NPs}.
The ideal cuboctahedron is a piece of the tetragonal FePt crystal ($L1_0$).
With the imposed periodic boundary conditions not all symmetry elements of the icosahedra 
are taken into account, and the relaxation has been performed with the lower point group $T_h$.
Nevertheless, due to large enough supercells all group elements of the $I_h$ symmetry,
including the fivefold rotation axis $C_5$, are preserved with the accuracy better than $10^{-3}$ \AA.
We consider the clusters with the magic numbers of atoms $N$ defined by the following formula:
\begin{equation}
N=(10n^3+15n^2+11n+3)/3=13, 55, 147, \ldots
\end{equation}
where $n$ is the number of closed shells. 
Additionally, the icosahedral particles without the central atom with N=12 and 54,
and the core-shell Fe$_{13}$Pt$_{42}$ system where studied.
For selected NPs, we have swapped Fe and Pt to investigate the influence of 
atomic configuration on structural and dynamical properties.
All studied systems are listed in Tab.~\ref{Tab1}.

\begin{table}[b!]
\caption{The list of studied nanoparticles with defined size, symmetry, and chemical composition.}
 \begin{tabular}{r l c c c c}
\hline
$N$ & System & Symmetry & Soft modes & Ground state  \\
\hline
13 & Fe$_5$Pt$_8$ & $D_{4h}$ & $A_{1g}$, $B_{1g}$ & $C_1$ \\
13 & Fe$_8$Pt$_5$ & $D_{4h}$ & $B_{2u}$, $B_{1g}$ & - \\
13 & FePt$_{12}$ & $I_h$ ($T_h$) & No &  $I_h$ ($T_h$)  \\
13 & Pt$_{13}$ & $I_h$ ($T_h$) & No &  $I_h$ ($T_h$)  \\
13 & Fe$_{12}$Pt & $I_h$ ($T_h$) & 2$T_u$, $E_u$, $A_g$, $T_g$, $A_u$ &  -\\
12 & Fe$_{12}$ & $I_h$ ($T_h$) & No & $I_h$ ($T_h$) \\
\hline
$55$ & Fe$_{24}$Pt$_{31}$ &  $D_{4h}$ & $A_{2u}$ & $C_{4v}$ \\
$55$ & Fe$_{31}$Pt$_{24}$ &   $D_{4h}$ & $B_{1g}$, $A_{2g}$, $E_u$ & - \\
$55$ & Fe$_{12}$Pt$_{43}$ &   $I_h$ ($T_h$) & $E_g$, $T_g$ &  -\\
$54$ & Fe$_{12}$Pt$_{42}$ &   $I_h$ ($T_h$) & No & $I_h$ ($T_h$) \\
$55$ & Fe$_{13}$Pt$_{42}$ &   $I_h$ ($T_h$) & No & $I_h$ ($T_h$) \\
$55$ & Fe$_{43}$Pt$_{12}$ &   $I_h$ ($T_h$) & $A_g$, $E_u$, $T_u$ & - \\
$54$ & Fe$_{42}$Pt$_{12}$ &   $I_h$ ($T_h$) & No & $I_h$ ($T_h$) \\
\hline
$147$ & Fe$_{67}$Pt$_{80}$ & $D_{4h}$ & No & $D_{4h}$ \\
$147$ & Fe$_{80}$Pt$_{67}$ & $D_{4h}$ & $A_{1u}$, $E_g$, $B_{1u}$, $B_{1g}$ & - \\
\hline
\end{tabular}
\label{Tab1}
\end{table}

Vibrational energies and polarization vectors were 
derived using the direct method \cite{Direct} implemented
in the PHONON program \cite{Phonon}.
First, we optimize the system to obtain the positions
of all atoms for the assumed symmetry.
Next, the Hellmann-Feynman (HF) forces are calculated
by displacing atoms from the equilibrium positions, one at a time,
in positive and negative directions by $0.03$~\AA.
The number of required displacements is determined by the
number of non-equivalent atoms and atom site symmetries.
From the HF forces obtained with respective displacements, the force-constants
are calculated and dynamical matrices are constructed. Finally, 
the vibrational energies and polarization vectors
are extracted by diagonalization of the dynamical matrix
at the $\Gamma$ point. 
The vibrational spectra of small NPs consist of separate lines, 
which are broadened using Gaussian functions with the width 1 meV FWHM 
for better presentation.

The obtained vibrational spectrum can be used to study the stability of the
system optimized in the assumed geometry. If all eigenvalues are real, 
the system is dynamically stable and there is no additional 
deformation of the particle. The presence of imaginary energies in a spectrum
indicates a possible structural deformation, which results in lowering 
of the symmetry. Polarization vectors of soft modes 
give complete information of atomic displacements 
from high-symmetry positions towards a more stable system
with the lower symmetry.

The described calculation procedure has two main advantages.
By using the higher symmetry of particles in the first step, 
the computational time can be reduced
comparing to calculations without symmetry constraints, and larger
systems can be studied in a reasonable time.
More important advantage is a possibility to reveal
the phonon modes, which are responsible for structural deformations
of particles. As we discuss below, such soft modes exist in some
NPs and lead to configurations with lower symmetries. 
For the stable systems, the vibrational heat capacities and entropies are calculated
within the harmonic approximation.

\section{Vibrational properties}

\subsection{Cuboctahedra}

\begin{figure}[t!]
\centering
\includegraphics[width=0.9\linewidth]{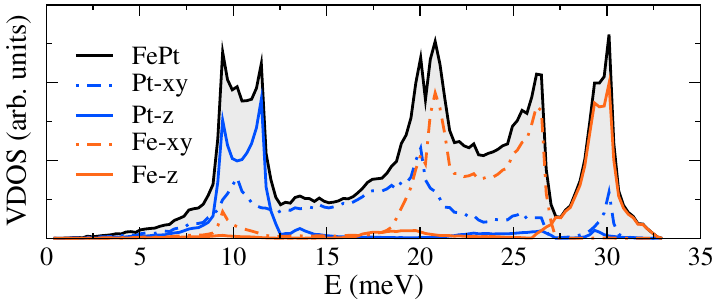}
\caption{Phonon density of states calculated for the FePt crystal in $P4/mmm$ symmetry ($L1_0$).}
\label{VDOSbulk}
\end{figure}

A cuboctahedral NP can be treated as a piece of the bulk, therefore,
the vibrational properties discussed in this section will be compared 
with the results obtained for the FePt crystal \cite{Sternik}. First, we recall 
the main features of the phonon DOS for this system presented in Fig.~\ref{VDOSbulk}.
The whole spectrum can be divided into two regions: low-energy part (below approx. 20~meV), 
where vibrations of Pt atoms dominate and high-energy region with the main contribution from Fe atoms.
The partial DOS functions projected along the $xy$ and $z$ directions demonstrate a large anisotropy
of lattice dynamics in this structure \cite{Sternik}. The lowest band (8-12~meV), consists mainly 
of movement of Pt atoms along the $z$ direction. Energies of Pt vibrations 
in $xy$ directions are distributed over almost the entire spectrum
with a maximum at 20~meV. Vibrations of Fe atoms mainly contribute to two energy regions:
18-28~meV ($xy$) and 26-33~meV ($z$).

\begin{figure}[b!]
\centering
\includegraphics[width=1.0\linewidth]{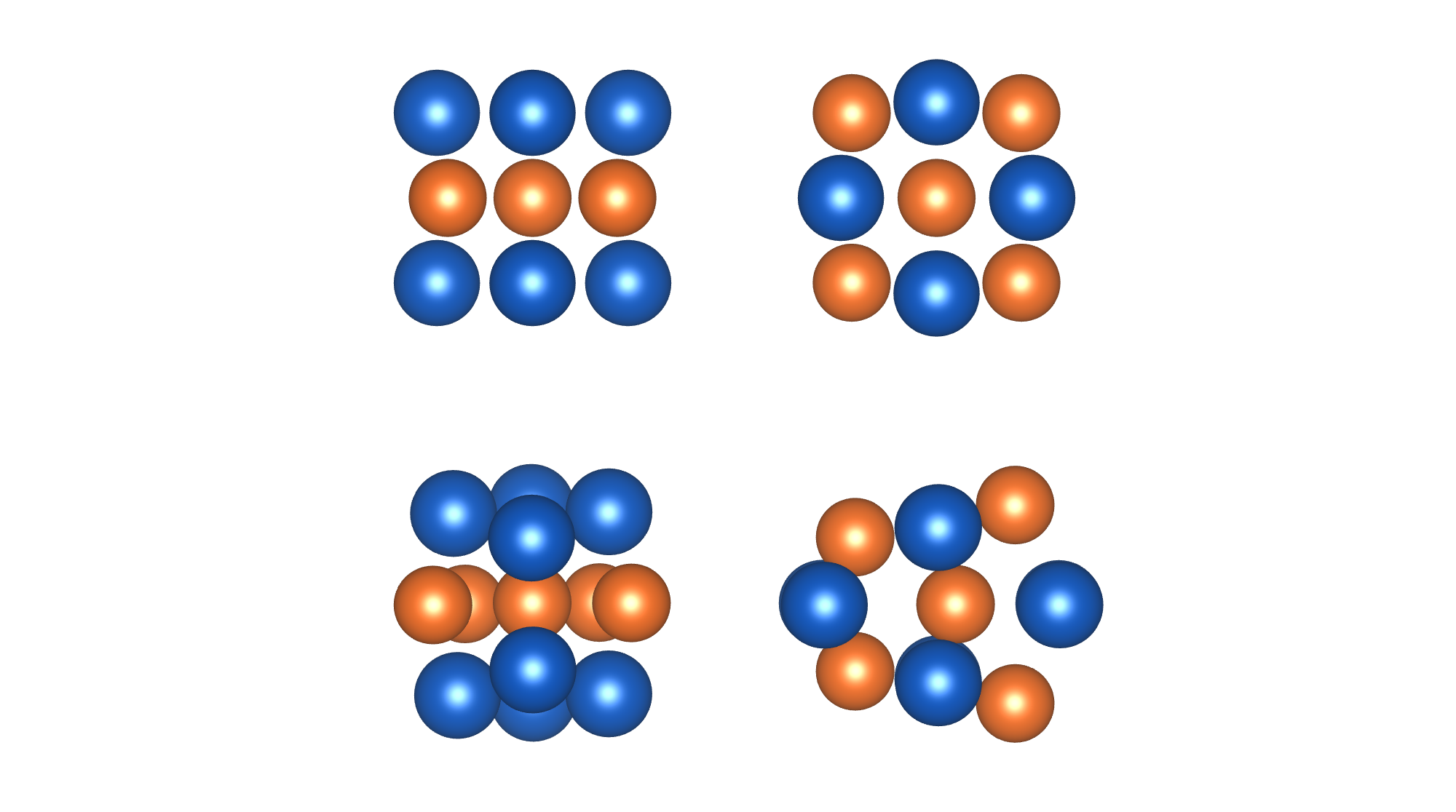}
\caption{Side (left column) and top view (right column) of the cuboctahedral Fe$_{5}$Pt$_{8}$ particle with $D_{4h}$ (top row) and $C_{1}$ symmetry (bottom row). 
Pt atoms are in blue (dark) and Fe atoms in red (light) colors.
The atomic radii were reduced accordingly to reveal displacements of atoms.}
\label{soft13}
\end{figure}

\begin{figure}[t!]
\centering
\includegraphics[width=0.9\linewidth]{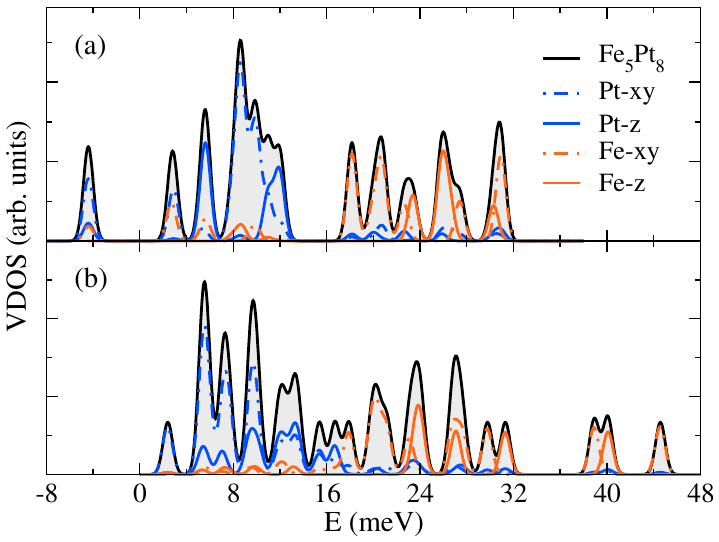}
\caption{Vibrational density of states of the Fe$_5$Pt$_8$ cuboctahedron calculated for (a) $D_{4h}$ and (b) $C_1$
structures.}
\label{VDOS13}
\end{figure}

${\bf Fe_5Pt_8 -}$ The smallest considered particle consists of 13 atoms arranged in
two Pt layers of four atoms each, separated by a single Fe layer of five atoms (see Fig.~\ref{soft13}).
This particle was optimized in the cubic supercell with size $d=20$~\AA,
which gives the distance between the particle and its nearest image
equal to 16~\AA.
The obtained distance between Fe and Pt layers equals 1.75~\AA,
so it is reduced with respect to the bulk value of 1.86~\AA.
Comparing to the bulk lattice parameter $a=3.85$~\AA, the corresponding
Fe-Fe and Pt-Pt distances are equal 3.48 and 3.92~\AA, respectively.
In the vibrational spectrum, shown in Fig.~\ref{VDOS13}, there are 33 modes ($3N-6$). 
The low-energy part, where Pt vibrations dominate, is separated by a gap (13-17~meV)
from the high-energy part of mainly Fe vibrations.
Performing calculations within the tetragonal $D_{4h}$ symmetry we found two soft modes with 
the $B_{1g}$ and $A_{1g}$ symmetries and the very close imaginary energies $4.45i$ meV and $4.39i$ meV, respectively.
Using the standard convention these imaginary modes are plotted with negative energies [see Fig.~\ref{VDOS13}(a)].
The fully symmetric mode $A_{1g}$ does not change the symmetry. 
In the $B_{1g}$ mode, two Pt atoms move outward and two inward (see the animation in Supplemental Material \cite{SM}),
distorting asymmetrically the particle and lowering its symmetry to $D_{2h}$.
The cluster optimized in this symmetry has lower total energy (by 32 meV/atom). However, this structure is still unstable
and after final optimization without symmetry elements ($C_1$), we get further lowering of the energy (by 15 meV/atom).
The final geometry of the particle is compared with the perfect cuboctahedron in Fig.~\ref{soft13}.
It has asymmetric shape, elongated in one direction and compressed in another. Some interatomic 
distances are strongly reduced, some of them enlarged.
The origin of such asymmetric deformation can be related with the changes in the electronic structure.
It removes the spin-down (minority) states from the Fermi energy and splits the spin-up states, partially removing its degeneracy,
similarly to the Jahn-Teller effect (see Supplemental Material \cite{SM}).  

The resulting vibrational spectrum consists of only real energies confirming the dynamic stability of the obtained
geometry [see Fig.~\ref{VDOS13}(b)]. Deformation of the particle causes large changes in the VDOS.
Symmetry lowering splits the degenerate modes and leads to effective broadening of the low- and high-energy bands
as well as closing the gap. More interesting is appearance of the vibrational states at much higher energies (above 38 meV),
even comparing to the phonon energies in the iron bcc crystal \cite{Lazewski}.
These vibrations originate from only one Fe atom located in the center of the particle, and
the highest mode at 44.6 meV corresponds to the in-plane vibration of this atom. 
Such high energies are induced by the Fe-Fe distance (2.36 \AA), strongly reduced in comparison
to the undistorted system (2.46 \AA). Also, the distances between the central atom and three Pt atoms
are reduced in the distorted particle.

${\bf Fe_8Pt_5 - }$ The nanoparticle Fe$_8$Pt$_5$ consists of one inner Pt layer with five atoms
and two surface Fe layers with four atoms each. The distance between nearest 
layers (1.86 \AA) equals to the bulk value. All interatomic distances
are reduced comparing to the Fe$_5$Pt$_8$ nanoparticle and the bulk FePt crystal.
This compressed geometry is very unstable and generates two soft 
modes with the symmetries $B_{1g}$ and $B_{2u}$ (see Tab.~\ref{Tab1}). 
The polarization vectors were used to distort the particle,
which relaxes to the lower symmetry ($C_{2v}$) with the reduced total energy 
by 38~meV/atom. Additional relaxation of the system without the symmetry
constraints leads to further lowering of the total energy by 6 meV/atom.
The final particle geometry is characterized by a strong deformation of the Fe layers 
and it is rather metastable, dynamically unstable configuration of atoms. 
There are still soft modes in the resulting vibrational spectrum (not shown).

\begin{figure}[t!]
\centering
\includegraphics[width=0.9\linewidth]{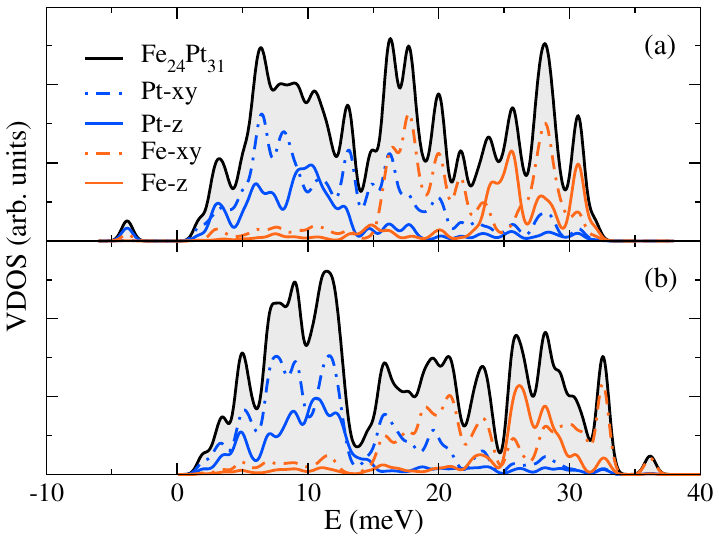}
\caption{VDOS of the Fe$_{24}$Pt$_{31}$ cuboctahedron calculated for (a) $D_{4h}$ and (b) $C_{4v}$ structures.}
\label{VDOS55}
\end{figure}

${\bf Fe_{24}Pt_{31} - }$ The next system we consider is the cuboctahedron with 55 atoms.
It is built of 3 Pt and 2 Fe layers, with slightly elongated outer interlayer distance (1.88~\AA)
and shortened inner distance (1.80~\AA) comparing to the bulk value (1.86~\AA).
The obtained total and partial vibrational spectra projected along the $xy$ and $z$ directions 
are presented in Fig.~\ref{VDOS55}. 
The differences between these projections show anisotropy in atomic vibrations.
Similarly to the phonon DOS of the FePt crystal, the high-energy region is dominated by Fe vibrations along the $z$ direction, 
and low-energy part by Pt atoms vibrating in the $xy$ plane.
The energy cut-off is only slightly shifted above the bulk limit (32 meV).
In contrast, the density of states largely increases in the range below 4 meV,
with main contribution from platinum surface atoms.  
Additionally, there is one soft mode $A_{2u}$ with the energy of $3.83i$~meV,
which reduces the symmetry $D_{4h}$ to its subgroup $C_{4v}$. 
In this soft mode, the Fe atoms move in $xy$ and Pt atoms in $z$ directions,
and only the surface atoms are involved (see Supplemental Material \cite{SM}).
After optimization of the system distorted by this mode, the total energy decreases by about 7 meV/atom only.
We verified also that additional optimization without symmetry constraints does not lower
the total energy. It means that the structure obtained within the $C_{4v}$ symmetry is the ground state
of the Fe$_{24}$Pt$_{31}$ cuboctahedron. 
\begin{figure}[t!]
\centering
\includegraphics[width=1\linewidth]{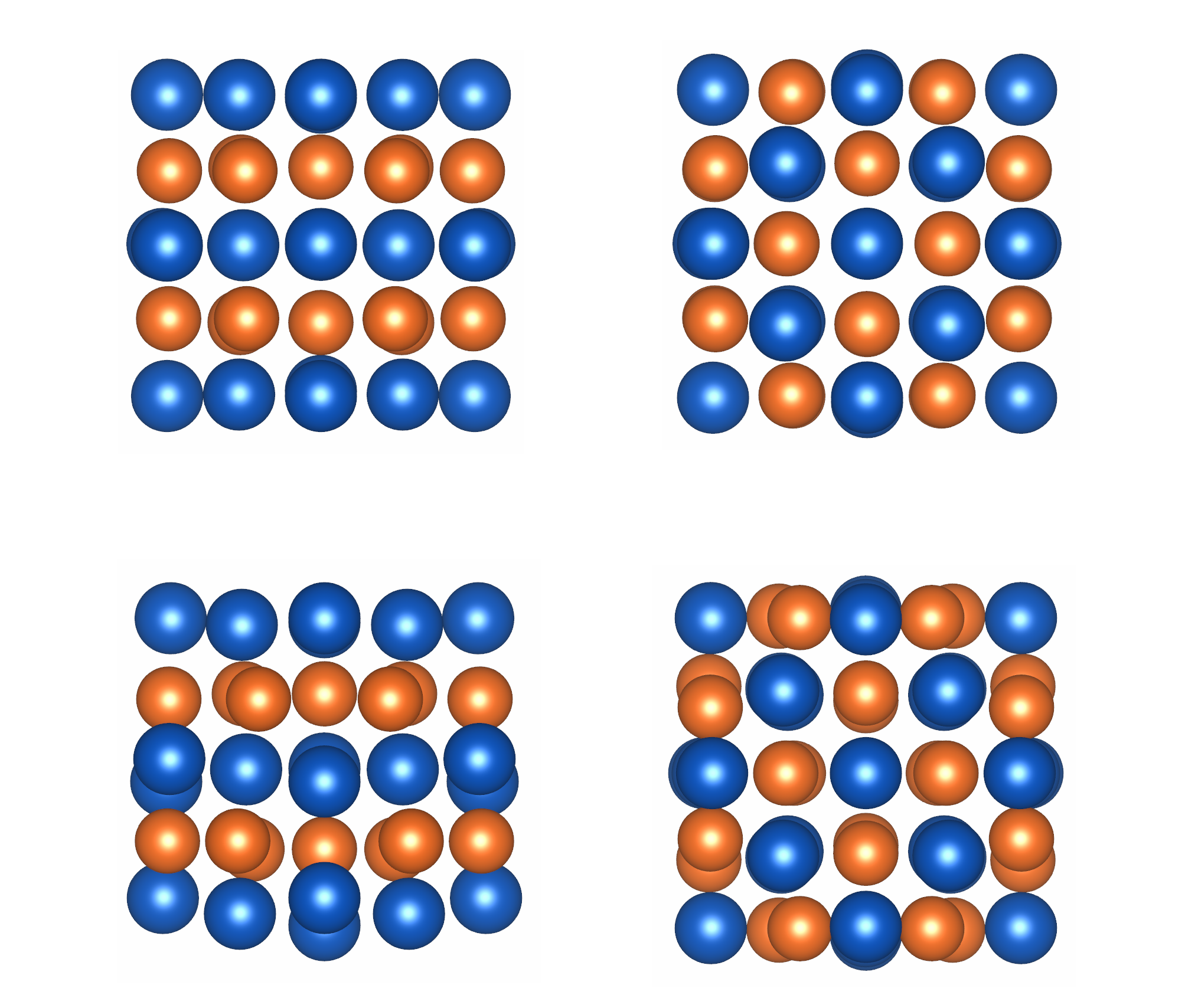}
\caption{Side (left column) and top view (right column) of the cuboctahedral Fe$_{24}$Pt$_{31}$ particle with $D_{4h}$ (top row) and $C_{4v}$ symmetry (bottom row). 
Pt atoms are in blue (dark) and Fe atoms in red (light) colors.}
\label{soft}
\end{figure}
The structures with the $D_{4h}$ and $C_{4v}$ symmetries are compared
in Fig.~\ref{soft}.
In the distorted structure, some Fe-Fe and Pt-Pt distances are reduced along the $xy$ 
and $z$ directions, respectively. Such dimerization was found for the first time in calculations
for larger FePt nanoparticles \cite{Gruner2}. Our study reveals the mechanism of deformation,
in which the soft mode brakes the $D_{4h}$ symmetry of perfect cuboctahedral particles.
The VDOS calculated for the distorted particle does not contain any soft mode [see Fig.~\ref{VDOS55}(b)],
however, there are substantial changes observed mainly at high energies. 
The energy cut-off is now shifted above 36~meV, and surprisingly, the highest mode
corresponds to the in-plane ($xy$) Fe vibrations.
 
${\bf Fe_{31}Pt_{24} - }$ We have studied also the system with the exchanged Fe and Pt atoms.
Having Fe atoms in the upper and bottom surface layers, this system is very
unstable. In the vibrational spectrum, there are three soft modes with the symmetries $B_{1g}$,
$A_{2g}$, and $E_u$, which all together strongly distort this NP lowering its symmetry to $C_1$.
After relaxation of the distorted particle, the total energy decreases by 51~meV/atom but 
there are still three soft modes and the system is dynamically unstable. 

\begin{figure}[t!]
\centering
\includegraphics[width=1.0\linewidth]{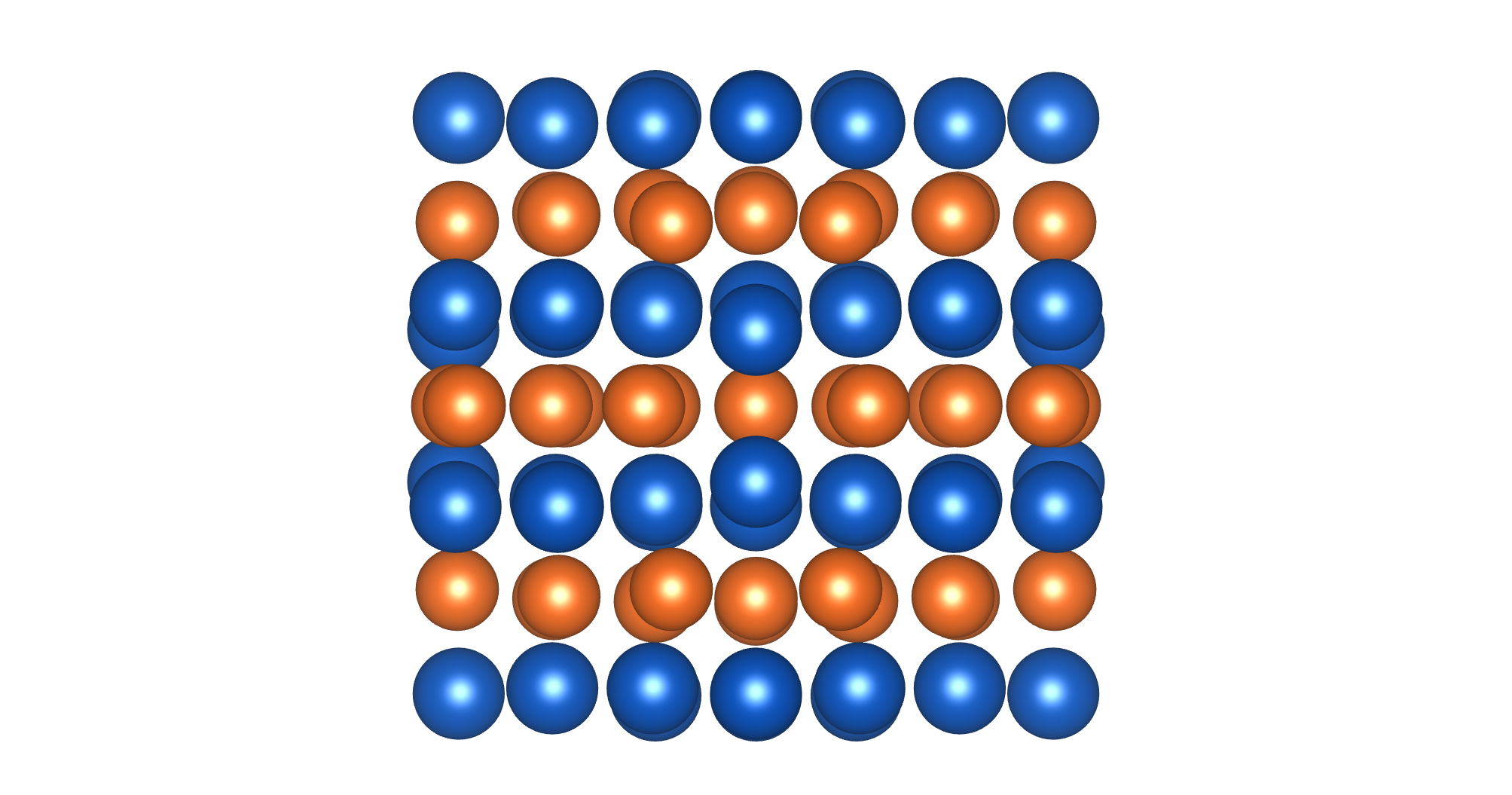}
\caption{Side view of the cuboctahedral Fe$_{67}$Pt$_{80}$ particle with $D_{4h}$ symmetry. 
Pt atoms are in blue (dark) and Fe atoms in red (light) colors.}
\label{dist147}
\end{figure}

\begin{figure}[h!]
\centering
\includegraphics[width=0.9\linewidth]{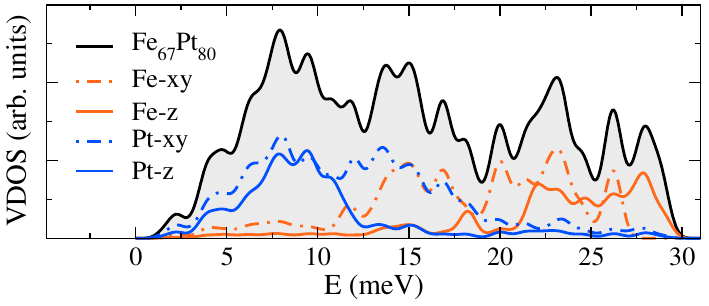}
\caption{VDOS of the Fe$_{67}$Pt$_{80}$ cuboctahedron.}
\label{VDOS147}
\end{figure}

${\bf Fe_{67}Pt_{80} - }$ The cuboctahedron with $N=147$ consists of 4 Pt and 3 Fe layers.
For this system, we have increased the linear size of the supercell to $d=25$~\AA,
to keep the distance between the particle and its images larger than 13~\AA.
The optimized particle is presented in Fig.~\ref{dist147}. 
Interestingly, it shows the same distortion (dimerization) as observed in Fig.~\ref{soft} 
and discussed in Ref.~[\onlinecite{Gruner2}]. 
In the obtained VDOS presented in Fig.~\ref{VDOS147}, there are no soft modes indicating that the structure
is dynamically stable. 
In this case, the $D_{4h}$ symmetry is not broken and the horizontal mirror plane is preserved. 
Thus, we can state a general rule that the FePt cuboctahedral NPs with an even number of Pt layers 
shows the distortion, which preserves the $D_{4h}$ symmetry and in those with an odd number of Pt layers, the symmetry is lowered 
by the soft-mode mechanism: $D_{4h}\rightarrow A_{2u} \rightarrow C_{4v}$.
The vibrational spectrum of this larger particle is more anisotropic than in the smaller one. 
The $z$-projected VDOS is shifted to the highest energies as in the bulk spectrum, however, the energy cut-off is smaller
than in the FePt crystal. Similarly to smaller particles, there is an increase of the low-energy
states coming from the Pt surface atoms.

${\bf Fe_{80}Pt_{67} - }$ Similarly to the smaller systems, the nanoparticle Fe$_{80}$Pt$_{67}$ with the Fe atoms in the surface
layers is also very unstable. There are four soft modes listed in Tab.~\ref{Tab1}. We have not studied
this nanoparticle any further.

\subsection{Icosahedra}

${\bf FePt_{12},  Fe_{12}Pt - }$ The smallest icosahedron consists of one central atom Fe (Pt) and twelve Pt (Fe) atoms placed
at equal distances from the center.
This distance is larger in the FePt$_{12}$ particle (2.59~\AA) than in the Fe$_{12}$Pt system (2.48~\AA).
However, both these distances are reduced comparing to the Fe-Pt distance in the bulk (2.70~\AA).
The NN distances in the shells are equal $d$(Pt-Pt)=2.72~\AA\ and $d$(Fe-Fe)=2.60~\AA, respectively.
The former one is reduced only by $2\%$ comparing to the distances between Pt atoms in the fcc crystal (2.77~\AA),
while the latter is larger by $4.4\%$ than between the Fe atoms in the bcc crystal (2.49~\AA).
As we show below, it has a strong impact on the stability of icosahedral particles. 

\begin{figure}[t!]
\centering
\includegraphics[width=0.9\linewidth]{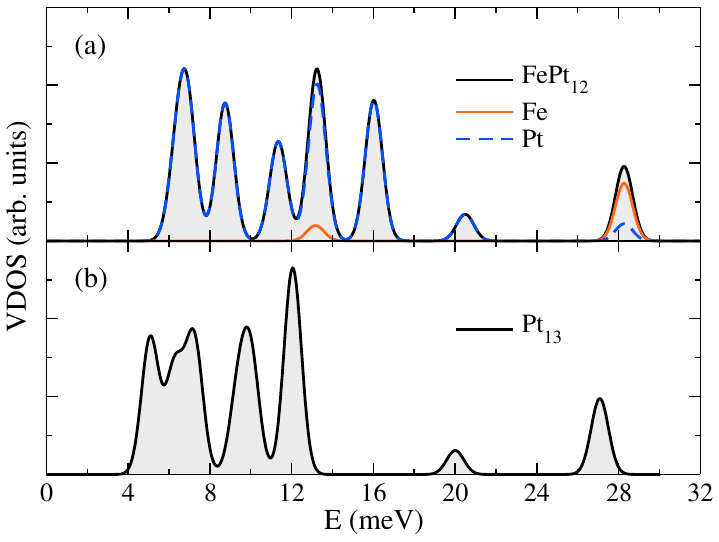}
\caption{Vibrational spectra of (a) FePt$_{12}$ and (b) Pt$_{13}$ icosahedra.}
\label{FePt12}
\end{figure}

The vibrational spectra of the smallest icosahedra are presented in Figs.~\ref{FePt12} and \ref{Fe12Pt}.
All these spectra are symmetric with respect to the $x$, $y$, $z$ components, 
therefore only the total and atom-projected DOS functions are presented.
For FePt$_{12}$, the spectrum consists of seven lines with a dominant contribution
from Pt atoms. The Fe atom takes part only in two modes at 13.3 and 28.3~meV.
The second highest mode at 20.5~meV is the fully symmetric $A_g$ mode in which all Pt atoms
vibrate along radial directions (breathing mode).
All energies are real indicating dynamical stability of this system.
This spectrum can be compared with the VDOS of the Pt$_{13}$ icosahedron
with all peaks shifted to lower energies [see Fig.~\ref{FePt12}(b)]. 
This shift is caused by larger interatomic Pt-Pt distances between the central atom
and the shell atoms (2.63~\AA) and between the atoms in the shell (2.77~\AA).
The latter equals to the NN distance in the bulk crystal.
The highest mode dominated by the movement of the central Pt atom decreases its energy to 27.1~meV,
while the breathing mode is shifted to 20~meV. The acoustic gap is reduced by about 2~meV.

In stark contrast, the spectrum of Fe$_{12}$Pt shows a few imaginary 
Fe modes in the range between $32i$ and $20i$~meV [Fig.~\ref{Fe12Pt}{a)]. They are induced by the overstretched
Fe-Fe distance, which strongly reduces the interatomic force constants,
and the small Fe-Pt distances.
The distortion induced by the soft modes removes the symmetry elements and lowers the total energy by 21~meV/atom.
The VDOS calculated without the symmetry constraints (with all atoms displaced) still shows a few
soft modes demonstrating that the particle cannot be completely stabilized in this geometry.
Replacing the central Pt atom with Fe does not stabilize the Fe$_{13}$ icosahedron (VDOS is not shown).
Although, the Fe-Fe distance in the shell (2.48~\AA) is comparable with the value in the bulk crystal, 
the distance between the central atom and the shell is strongly reduced (2.36~\AA).
Relaxation of the Fe$_{13}$ icosahedron without the symmetry lowers the total energy by 63~meV/atom,
however, the VDOS calculated using displacements of all atoms still exhibits the imaginary modes.
As shown by numerous test calculations, Fe$_{12}$Pt or Fe$_{13}$ icosahedra can be stabilized 
only by removing the central Pt or Fe atom and relaxing the positions of all the remaining Fe atoms. 
After relaxation, the radius of the Fe shell decreases to 2.28 \AA,
and the Fe-Fe distance to 2.39~\AA. It results in the shift of the lowest mode to 11.4~meV
largely increasing the acoustic gap, and the shift of the highest mode to the lower energy 
34.3~meV [see Fig.~\ref{Fe12Pt}(b)].

${\bf Fe_{12}Pt_{43} - }$The icosahedral particles with $N=55$ consists of a central atom and two shells with 12 and 42 atoms.
In Fe$_{12}$Pt$_{43}$, the radius of the Fe shell equals to 2.59~\AA\ and is larger than in Fe$_{12}$Pt
due to the attraction from the Pt shell.
In this outer shell, the Pt atoms are located in two distinct distances from the center: 5.05~\AA{} (corner atoms)
and 4.48~\AA{} (edge atoms). The NN distances in the shells are $d$(Fe-Fe)=2.72~\AA\ and $d$(Pt-Pt)=2.66~\AA.
The VDOS of this particle is presented in Fig.~\ref{ico55-PFP}(a).
There are two soft modes with very close energies at $16.35i$ meV ($E_g$) and $16.33i$ meV ($T_g$) meV of mainly Fe character. 
The origin of this soft mode is similar as in the Fe$_{12}$Pt particle: enlarged Fe-Fe distance in the Fe shell and reduced Fe-Pt distance.
The energy vibrations of the outer Pt shell have a dominant contribution in the range between 4 and 24 meV
with some participation of Fe atoms. In this energy range, the central Pt atom contributes only to three modes. 
Two highest modes of mainly Fe character are separated by about 4 meV and are located at $\sim32$ and $\sim36$ meV.
Interestingly, there is also small contribution of the central Pt atom in the highest mode.

\begin{figure}[t!]
\centering
\includegraphics[width=0.9\linewidth]{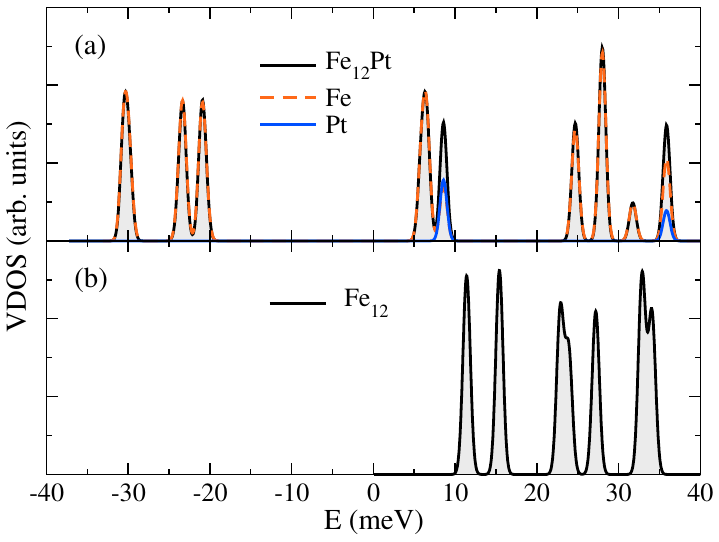}
\caption{Vibrational spectra of (a) Fe$_{12}$Pt and (b) Fe$_{12}$ icosahedra.}
\label{Fe12Pt}
\end{figure}

We have used the soft modes to distort the particle and relax it to the lower symmetry $S_2$. The relaxation decreases the particle energy only by 1~meV/atom and does not remove completely the soft modes from the VDOS.
As in the smaller icosahedron, the particle can be stabilized by removing the central Pt atom,
which leads to the two-shell system Fe$_{12}$Pt$_{42}$ with modified interatomic distances.
All radial distances decrease; the radius of the Fe shell contracts to 2.48~\AA\ and the distances of the Pt atoms from the center
are reduced to 4.99 and 4.46~\AA.
These changes not only stabilize the structure (removing the soft mode) but modify the whole spectrum of vibrations. 
Due to smaller Pt-Pt distances, the spectrum of Pt vibrations is shifted to higher energies. 
At the same time, the energies of two highest modes of mainly Fe character decrease to 29 and 32.6~meV due to larger Fe-Pt distances,
thus the gap separating these modes is substantially reduced. The highest mode is the internal breathing mode of the Fe shell.

\begin{figure}[t!]
\centering
\includegraphics[width=0.9\linewidth]{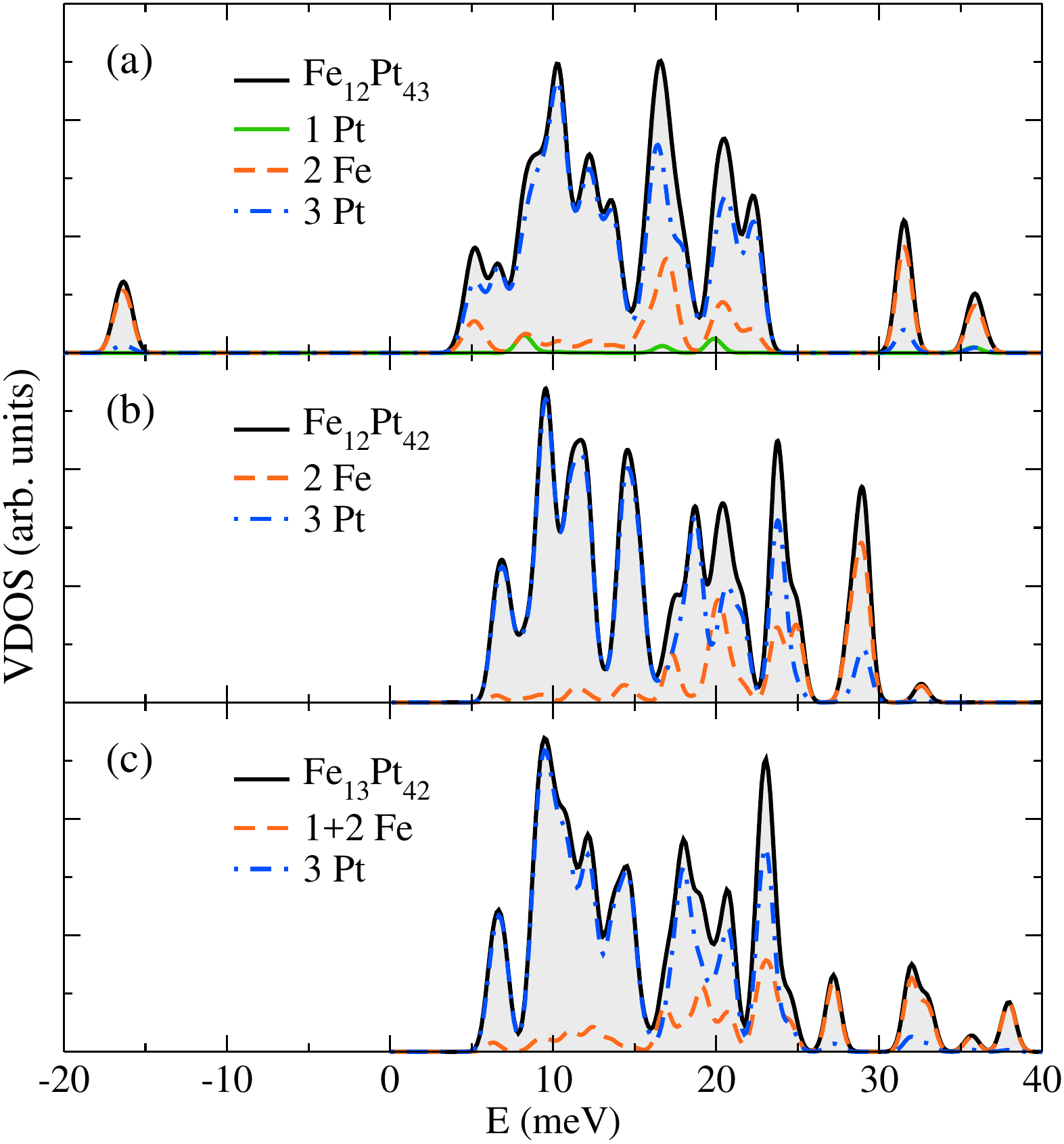}
\caption{Vibrational density of states of (a) Fe$_{12}$Pt$_{43}$, (b) Fe$_{12}$Pt$_{42}$, and (c) Fe$_{13}$Pt$_{42}$ icosahedra.}
\label{ico55-PFP}
\end{figure}

${\bf Fe_{13}Pt_{42} - }$We have performed calculations for the similar core-shell system Fe$_{13}$Pt$_{42}$ 
with the central Pt atom replaced by the Fe atom. Due to the smaller atom in the centre, all interatomic distances are
reduced comparing to the Fe$_{12}$Pt$_{43}$ system. The larger change is observed for the Fe core. The radius contracts
to 2.53~\AA\ and the NN Fe-Fe distance to 2.66~\AA. The NN distance in the Pt shell (2.64~\AA) and the distances
from the centre to corner (5.02~\AA) and edge (4.47~\AA) Pt atoms are only slightly reduced and very similar to those
in the NP without the central atom. 
The VDOS presented in Fig.~\ref{ico55-PFP}(c) shows the dynamical stability of this system. Comparing to the Fe$_{12}$Pt$_{42}$ system, the energies
of Pt vibrations are slightly extended to higher energies. The spectrum of Fe vibrations shows larger changes and the highest
peak is shifted to 38~meV, above the values found in both similar systems.

${\bf Fe_{43}Pt_{12} - }$The VDOS of the Fe$_{43}$Pt$_{12}$ system is presented in Fig.~\ref{ico55-FPF}(a).
The optimized radial distances are equal to 2.64~\AA\ for the Pt shell, and 5.09 and 4.37 \AA\ for the Fe shell.
The NN distances in the Pt and Fe shells equal 2.78 and 2.68~\AA, respectively.
Strongly elongated Fe-Fe distance leads to three unstable Fe modes at $3.45i$~meV ($A_g$), $2.27i$~meV ($E_u$), and $2.09i$~meV ($T_u$).
The vibrations of Fe and Pt atoms are distributed over the whole energy spectrum (2-34 meV) with
the dominant contribution from the Fe surface shell. 
The central Fe atom is involved only in the highest modes at about 32~meV.
We have repeated calculations without the central Fe atom obtaining the stable structure
with all real vibrational energies presented in Fig.~\ref{ico55-FPF}(b).
The radii of both shells are slightly reduced to 2.60~\AA (Fe) and 5.05-4.36~\AA (Pt),
which results in shortening of the NN distances $d$(Pt-Pt)=2.74 and $d$(Fe-Fe)=2.65~\AA.
It stabilizes the particle and shifts the whole spectrum of vibrations to higher energies.
The lowest mode at 5.3~meV, which defines the acoustic gap, and the highest mode at 33.6~meV 
are mainly of the Fe character.

\begin{figure}[t!]
\centering
\includegraphics[width=0.9\linewidth]{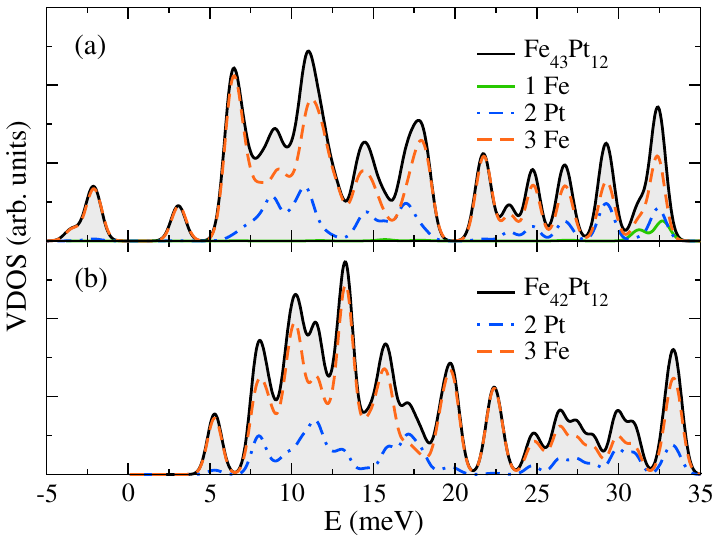}
\caption{Vibrational density of states of (a) Fe$_{43}$Pt$_{12}$ and (b) Fe$_{42}$Pt$_{12}$ icosahedra.}
\label{ico55-FPF}
\end{figure} 
 
\subsection{Thermodynamic properties}

For the stable nanoalloys, the thermodynamic properties can be studied within the harmonic approximation.
In Figs.~\ref{hc1} and \ref{hc2}, we have compared the vibrational heat capacity calculated for the NPs
with different geometries and number of atoms in the temperature range $T=0-300$ K. 
The insets present the results below $T=30$ K.
At low temperatures, the heat capacity of cuboctahedral particles presented in Fig.~\ref{hc1} decreases 
with the increasing number of atoms. It is connected with the relatively larger VDOS at lower energies 
in the smallest particles. This tendency is reversed around $T=50$~K. Since in the smaller particles,
the energy cut-off is shifted to larger values, their heat capacity is reduced at higher temperatures 
comparing to larger particles. At each temperature, all systems have larger heat capacity 
than the bulk FePt crystal.

\begin{figure}[t!]
\centering
\includegraphics[width=0.9\linewidth]{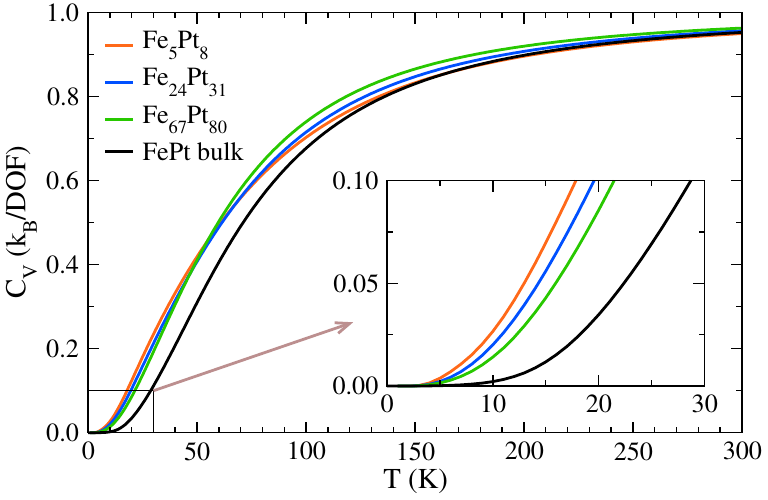}
\caption{Vibrational heat capacity for the cuboctahedral particles compared with the FePt crystal 
lattice heat capacity calculated per one degree of freedom (DOF).}
\label{hc1}
\end{figure}

In Fig.~\ref{hc2}, we have compared the heat capacity of two icosahedral particles Fe$_{12}$Pt$_{42}$
and Fe$_{42}$Pt$_{12}$ with one of the cuboctahedra Fe$_{24}$Pt$_{31}$. They have similar
sizes and total numbers of atoms. At the lowest temperatures, the cuboctahedron has larger heat
capacity than the icosahedra because of the larger acoustic gap in the icosahedra (5-6 meV)
than in the cuboctahedron (1 meV). Above around 60 K, the heat capacity of the icosahedra exceeds
the values for the cuboctahedron, which exhibits the higher cut-off of the vibrational energies.
Interestingly, in spite of different concentration of Fe and Pt atoms, both icosahedral NPs
have very similar heat capacities in the entire range of temperatures. It demonstrates
that the thermodynamic properties of nanoalloys are to a large extend determined by the geometry.
At the lowest temperatures, due to the smaller acoustic gap, the heat capacity in Fe$_{42}$Pt$_{12}$ 
is slightly larger than in Fe$_{12}$Pt$_{42}$. 

In Table~\ref{Tab2}, the vibrational entropy calculated for the selected nanoparticles
is compared with the entropy of the FePt crystal.
The average entropy is very similar for all nanoparticles and it is approximately $20\%$ larger than in the bulk.
The contribution from Fe vibrations in the subsurface layers or in the particle's core is slightly enhanced (about $6 \%$),
but it is notably increased by $44\%$ in the surface shell of the Fe$_{42}$Pt$_{12}$ particle.
It is consistent with the unstable character and low melting temperatures of particles with Fe surface layers \cite{Jochym}.
The entropy of Pt vibrations in the icosahedral NPs is very similar in the core and surface layers and it is larger 
by $7 \%$ comparing to the bulk. The larger increase by about $23 \%$ is found for the surface of the cuboctahedral Fe$_{24}$Pt$_{31}$ particle.
It shows that cuboctahedral particles are less stable than the icosahedra with the Pt outer shells.

\begin{figure}[t!]
\centering
\includegraphics[width=0.9\linewidth]{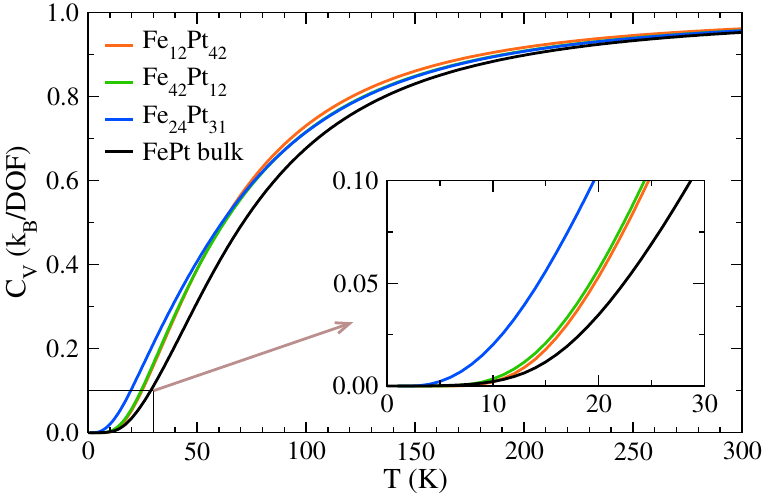}
\caption{Comparison of the vibrational heat capacity of two icosahedra with the cuboctahedral particle and the FePt crystal 
lattice heat capacity calculated per one degree of freedom (DOF).}
\label{hc2}
\end{figure}

\begin{table}[h!]
\caption{Vibrational entropy calculated in units $k_B$ per one degree of freedom (DOF).}
 \begin{tabular}{c c c c c c}
\hline
 & FePt($L1_0$) & Fe$_{24}$Pt$_{31}$ & Fe$_{12}$Pt$_{42}$ & Fe$_{13}$Pt$_{42}$ & Fe$_{42}$Pt$_{12}$ \\
\hline
Total & 1.376 & 1.663 & 1.632 & 1.609 &  1.666   \\
Fe & 1.144 &  1.246 & 1.204 & 1.216 & 1.646 \\
Pt & 1.608 &  1.986 & 1.754 & 1.731 & 1.736 \\
\hline
\end{tabular}
\label{Tab2}
\end{table}

\section{Discussion and summary} 

The results obtained for all studied FePt systems have been summarized in Table \ref{Tab1}.
The cuboctahedral NPs with the odd number of Pt layers stabilize in the distorted structure 
with the lower symmetry induced by the soft mode. In the systems with the even number of Pt layers such distortion
is observed without breaking the $D_{4h}$ symmetry.
In the case of icosahedra with the perfect shell geometry, only the smallest particle FePt$_{12}$ is dynamically stable.
Other studied icosahedral systems are unstable due to the stretched Fe-Fe and reduced Fe-Pt distances
and they can be stabilized by removing a central Fe or Pt atom or by replacing platinum by iron 
in the Fe$_{12}$Pt$_{43}$ nanoparticle. 
This result agrees with the previous observations that a large atom in the center of an icosahedron
induces a large compressive stress and can destabilize the structure \cite{Bochicchio}.
The single-element icosahedra can be more stable without the central atoms \cite{Mottet97},
but in the core-shell particles with smaller atoms in the core, the central vacancy is not favored \cite{Laasonen}.

Both mechanisms of stabilization discussed here induce changes not only at the lowest vibrational energies but also 
at higher energies due to the modified interatomic distances.
In particular, the deformations of the cuboctahedra with N=13 and 55 induced by the soft modes
lead to the appearance of the Fe modes with the energies higher than in the bulk
FePt alloy \cite{Sternik}. 

All these calculations are performed for the ground states corresponding
to $T = 0$ K. Thermal fluctuations may stabilize some of the considered systems, which exhibit
the soft-mode behavior. Indeed, the {\it ab initio} MD studies demonstrate very
high stability of the icosahedral particles with the Pt atoms in the outer shell \cite{Jochym}.
In these simulations performed at various temperatures, the Fe$_{12}$Pt$_{43}$ particle
is very stable even at $T=1000$ K and new preliminary studies indicate a very high melting temperature
(around 1500 K). It shows that the unstable behavior due to the presence of central atoms,
may be relevant only at very low temperatures.  

The MD simulations show that the particles with the Fe atoms in the outer shell are
very unstable and strongly distorted even at low temperatures \cite{Jochym}.
It is induced by a tendency of Pt atoms to move to the surface layer even in the platinum
deficient particles.
This is in perfect correspondence with the present study. Each of the three cuboctahedral NPs 
with the Fe atoms in the surface layers exhibits a few soft modes, which strongly distort their geometries.
For none of these systems, we could obtain the dynamically stable structure and manage to calculate the vibrational
spectrum. Two icosahedral particles with the Fe atoms in the outer shell were stabilized
only by removing the central atom. 

Similarly to other systems \cite{Rojas,Shafai,Carles}, the modified vibrational spectra of FePt nanoparticles influence their thermodynamic properties.
The calculated heat capacities of nanoparticles are increased comparing to the lattice
heat capacity of the crystal due to the enhanced VDOS at lower energies.
At low temperatures, the heat capacity of cuboctahedral particles is larger than in icosahedra, which have larger acoustic gaps.
Also, the vibrational entropy of nanoparticles is enhanced comparing to the bulk material.
The largest increase of entropy is found for the Fe atoms in the outer shell of icosahedron,
which confirms the unstable character of this system.

\begin{acknowledgments}

The authors acknowledge support by the COST Action MP0903 
"Nanoalloys as Advanced Materials: From Structure to Properties and Applications" 
and by the Polish National Science Center (NCN) under Project No. 2011/01/M/ST3/00738.

\end{acknowledgments}

\end{document}